\batchmode
\makeatletter
\def\input@path{{\string"C:/Users/braive-adm/Documents/OptoMechanics/Article/Vibrational Res/Version5/\string"}}
\makeatother
\documentclass[10pt,twocolumn,english]{paper}
\usepackage[latin9]{inputenc}
\usepackage{xcolor}
\usepackage{pdfcolmk}
\usepackage{amsmath}
\usepackage{amsthm}
\usepackage{amssymb}
\usepackage{graphicx}
\usepackage[authoryear]{natbib}
\PassOptionsToPackage{normalem}{ulem}
\usepackage{ulem}

\makeatletter

\DeclareTextSymbolDefault{\textquotedbl}{T1}
\providecolor{lyxadded}{rgb}{0,0,1}
\providecolor{lyxdeleted}{rgb}{1,0,0}

\DeclareRobustCommand{\lyxsout}[1]{\ifx\\#1\else\sout{#1}\fi}

\numberwithin{equation}{section}
\numberwithin{figure}{section}
\newcommand{\lyxaddress}[1]{
	\par {\raggedright #1
	\vspace{1.4em}
	\noindent\par}
}

\usepackage{babel}

\makeatother

\usepackage{babel}
\begin{document}
\title{Weak signal enhancement by non-linear resonance control in a forced
nano-electromechanical resonator}
\author{Avishek Chowdhury$^{1}$, Marcel G. Clerc$^{2}$, Sylvain Barbay$^{1}$,
Isabelle Robert-Philip$^{3}$and Remy Braive$^{1,4}$}
\maketitle

\lyxaddress{$^{1}$ Centre de Nanosciences et de Nanotechnologies, CNRS, Univ.
Paris-Sud, Universit� Paris-Saclay, Palaiseau, France}

\lyxaddress{$^{2}$ Departamento de F\textcolor{black}{�sica and Millennium Institute
for Research in Optics,} \textcolor{black}{Facultad de Ciencias F�sicas
y Matem�ticas, Universidad de Chile, Casilla 487-3, Santiago, Chile}}

\lyxaddress{$^{3}$ Laboratoire Charles Coulomb, Universit� de Montpellier,CNRS,
34000 Montpellier, France}

\lyxaddress{$^{4}$ Universit� de Paris, 75207 Paris Cedex 13, France}
\begin{abstract}
Driven non-linear resonators can display sharp resonances or even
multistable behaviours amenable to induce strong enhancements of weak
signals. Such enhancements can make use of the phenomenon of vibrational
resonance whereby a weak low-frequency signal applied to a bistable
resonator can be amplified by driving the non-linear oscillator with
another appropriately-adjusted non-resonant high-frequency field.
Here we demonstrate the resonant enhancement of a weak signal by use
of a vibrational force yet in a monostable system consisting of a
driven nano-electromechanical nonlinear resonator. The oscillator
is subjected to a strong quasi-resonant drive and to two additional
tones: a weak signal at lower frequency and a non-resonant driving
at an intermediate frequency. We show experimentally and theoretically
a significant enhancement of the weak signal thanks to the occurence
of vibrational resonance enabled by the presence of the intermediate
frequency driving. We analyse this phenomenon in terms of coherent
nonlinear resonance manipulation. Our results illustrate a general
mechanism which may have applications in the fields of radio-frequency
signal processing or sensing for instance.
\end{abstract}

\section*{Introduction}

\noindent In bistable systems, weak periodic signals can be amplified
by use of external driving. Such external driving can be some noise
of appropriate strength in the case of stochastic resonance \citet{BenziJPAMG81},
or a high-frequency harmonic signal of appropriate amplitude in the
case of vibrational resonance \citet{LandaJoPAMaG00}. Both physical
phenomena share qualitative features including a resonant-like behaviour,
though the underlying mechanisms differ. Time matching criterion dependent
on the applied noise amplitude required for stochastic resonance is
replaced, in the case of vibrational resonance, by an amplitude criterion
equivalently to a parametric amplification near the critical point.
Both phenomena have been reported in many different areas including
electronics \citet{ZaikinPRE02,FauvePLA83}, optics \citet{McNamaraPRL88,BarbayPRE00,ChizhevskyPRL03,ChizhevskyPRA05}
or neurobiology \citet{UllnerPLA03,DouglassN93}. In nanomechanics,
the bistable system is usually a simple nonlinear resonator and bistability
arises thanks to a quasi-resonant forcing. The oscillator then features
two equilibrium states of different amplitudes and phases for the
same values of parameters. In this regime, substantial resonant enhancement
of a weak and slowly modulated signal through stochastic resonance
can be achieved either by use of amplitude \citet{BadzeyNat05,GuerraNL09,VenstraNatCom13}
or phase \citet{ChowdhuryPRL17} noise. When the external driving
is no more stochastic but rather a harmonic signal of high frequency,
a little bit of care has to be taken. The system is then subjected
to forces occurring on three different timescales: the one of the
signal, the one of the external drive and the one of the forcing.
In the standard picture of vibrational resonance, the signal must
have a much smaller frequency than the one of the external drive.
We here show that the occurence of vibrational resonance in a forced
system requires the external driving frequency not only to be higher
than the signal frequency but also to be lower than the forcing frequency.
Most importantly, we show in that case that the high-frequency driving
amplitude renormalises the forced nonlinear resonator response through
the manipulation of the nonlinear resonance. We argue that this effect
could be used besides the one we are presenting here, for another
purpose using pervasive biharmonic signals such as inertial Brownian
motors \citet{MACHURA2010445} and the Global Positioning System \citet{Kaplan05},
and illustrates thus a general mechanism for nonlinear resonance manipulation.

\section*{Results}

\begin{figure*}
\begin{centering}
\includegraphics[width=1\textwidth]{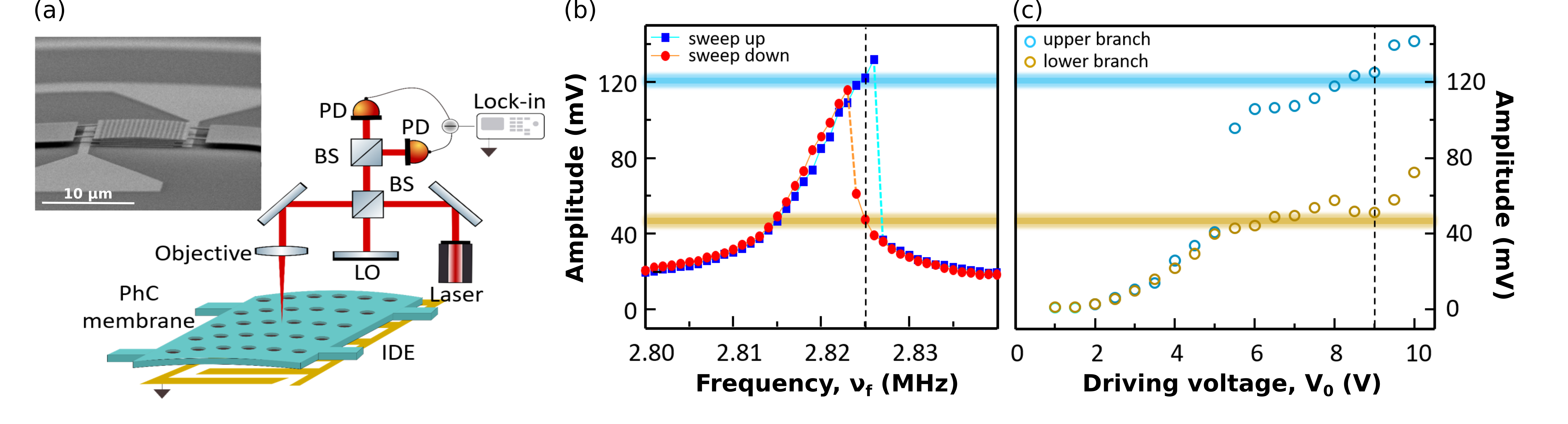}
\par\end{centering}
\caption{(a) Schematic of the experimental set-up used for the actuation and
detection of the mechanical motion of the membrane (PhC: photonic
crystal; IDE: Interdigitated electrodes; LO: local oscillator; BS:
beamsplitter; PD: photodiode). Inset: Scanning electron microscopic
image of the resonator consisting of a suspended InP membrane with
a thickness of 260 nm and a surface of 20$\times$10 $\mu$m$^{2}$
and integrating gold interdigitated electrodes underneath at a 400
nm distance. (b) Amplitude of the forced mechanical fundamental mode
as a function of the forcing frequency $\nu_{\mathrm{f}}$ in a sweep
up (blue dots) and sweep down (red dots) experiment for $V_{0}=9V$.
The vertical dashed line lies at a frequency of $2.825\,MHz$. (c)
Amplitude of the forced mechanical fundamental mode as a function
of the amplitude of the applied voltage $V_{0}$ for a forcing frequency
fixed at $\nu_{\mathrm{f}}=2.825\,MHz$. The vertical dashed line
lies at a voltage amplitude of $9V$.}
\label{Fig1}
\end{figure*}

In our experiments, the resonator consists of a non-linear nano-electromechanical
oscillator formed by a thin micron-scale InP suspended membrane (see
Fig. \ref{Fig1}-a). The membrane's out-of-plane motion is induced
by applying an AC voltage $V(t)$ on integrated metallic interdigitated
electrodes placed underneath the membrane at a sub-micron distance
(see Figure \ref{Fig1}-a) (See \citet{ChowdhuryAPL16} and Methods
for more details). It is placed in a low-pressure chamber ($10^{-4}\,mbar$)
in order to reduce mechanical damping. The out-of-plane motion of
the membrane is probed optically (see Fig. \ref{Fig1}-b) thanks to
a Michelson interferometer whose one end mirror is formed by the oscillating
membrane (see Fig. \ref{Fig1}-a). The membrane mechanical fundamental
mode of oscillation lies at 2.82 MHz in the linear regime with a mechanical
quality factor of $Q_{M}\sim10^{3}$.

The AC forcing voltage lies in the MHz regime and writes : 
\begin{equation}
V(t)=V_{0}cos[2\pi\nu_{\mathrm{f}}t]
\end{equation}
Here $V_{0}$ is the amplitude of the applied voltage while $\nu_{\mathrm{f}}$
denotes the frequency of the quasi-resonant forcing. When sweeping
up and down the frequency $\nu_{\mathrm{f}}$ in the vicinity of the
fundamental mode frequency, asymmetry in the mechanical response spectrum
appears for a sufficiently high driving amplitude $V_{0}>5.5V$ (see
Fig. \ref{Fig1}-b). Hysteresis behaviour becomes prominent and two
stable points, represented by the low and the high amplitude values
of the mechanical motion, co-exist. The evolution of the bistable
region width as a function of $V_{0}$ is shown in Fig. \ref{Fig1}-b
Right. The closing of this bistable region for increasing $V_{0}$
cannot be observed due to limited voltage handled at the electrodes
terminals. In the following, $V_{0}$ will be set to $9V$ in order
to be deeply in the bistable regime, and the forcing frequency is
set at $\nu_{\mathrm{f}}=2.825\,MHz$, close to the middle of the
hysteresis region, in order to get symmetrical potentials \citet{ChowdhuryPRL17}.

\subsection*{Externally induced dynamics in the time domain}

\begin{figure}
\centering{}\includegraphics[width=1\columnwidth]{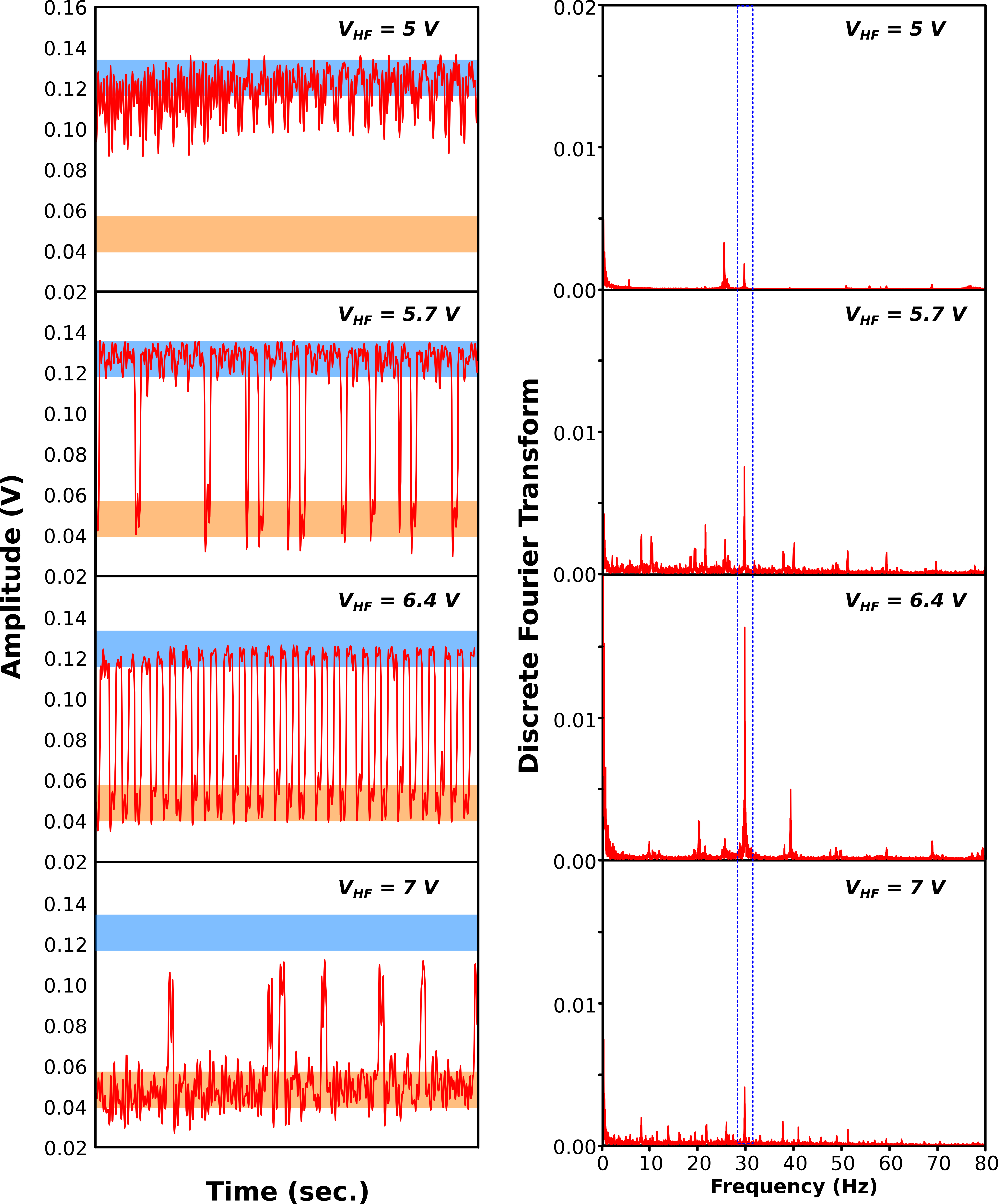}\caption{(Left) Time series of the mechanical mode amplitude for a weak modulation
with $\gamma=0.1$, $\nu_{\mathrm{m}}=30\,Hz$ and for increasing
high-frequency signal intensities $V_{HF}=5\,V,5.7\,V,6.4\,V$ \&
$7.0\,V$ from top to bottom. The amplitude of the two stable states
corresponds to the blue and orange-shaded regions on each graph. (Right)
Discrete Fourier Transform of the time series displayed on the left.
The vertical shaded lines enclose the modulation frequency $\nu_{m}$
of the weak signal.}
\label{Fig2}
\end{figure}

Jumps between the two stable states of oscillation can be induced
by slowly modulating the forcing amplitude. This scenario can be implemented
by applying to the electrodes a voltage in the form of: 
\begin{equation}
V(t)=V_{0}(1+\gamma\cos(2\pi\nu_{\mathrm{m}}t))\cdot cos[2\pi\nu_{\mathrm{f}}t]
\end{equation}
where $\gamma$ and $\nu_{\mathrm{m}}$ denote respectively the modulation
index and the frequency of the amplitude modulated signal. Yet, a
sufficiently high modulation amplitude is needed to drive the system
in order to overcome the barrier height and to induce inter-well motion
following the applied modulation. In the case of a weak amplitude
modulated signal (as in our experiments with $\gamma$ set at $0.1$),
the system is solely subjected to intra-well modulated motion as can
be seen in Fig. \ref{Fig2} top for the resonator being initially
prepared in its upper state. Amplification of the weak modulated signal
following jumps of the system between the two states can however still
be induced in that case by adding an external driving with a frequency
that is much higher than the frequency of the weak modulation, but
still lower than the forcing frequency.

\noindent 

In this scenario reproduced in our experiment, the external drive
takes the form of an additive amplitude modulation voltage of amplitude
$V_{HF}\equiv\delta*V_{0}$ and frequency $\nu_{HF}$ such that $\nu_{\mathrm{m}}\ll\nu_{\mathrm{HF}}\ll\nu_{\mathrm{f}}$.
The total applied voltage then writes: 
\begin{equation}
V(t)=V_{0}\left[1+\gamma\cos(2\pi\nu_{\mathrm{m}}t)+\delta\cos(2\pi\nu_{\mathrm{HF}}t)\right]\cos\left(2\pi\nu_{\mathrm{f}}t\right)
\end{equation}

\noindent Figure \ref{Fig2} shows time series of the mechanical motion
amplitude for $\nu_{m}=30\,Hz$ and increasing amplitudes of the amplitude
modulation at high frequency $\nu_{\mathrm{HF}}=200\,kHz>6500\cdot\nu_{\mathrm{m}}$.
The system starts in its upper state (high amplitude state) where
the small signal modulation is visible as a small intra-well motion.
As the amplitude of the external driving increases, switching events
between the two stable states become more prevalent. At first, occasional
inter-well transitions occur, weakly locked to the modulation signal.
For $V_{HF}=6.4\,V$, the system response gets completely synchronised
with the applied weak and low frequency modulation. Further increase
of the additional external drive amplitude worsens the synchronisation
and the system drops into its lower amplitude state, where a small
intra-well modulation is visible. There is thus an optimal amplitude
of the external drive which maximises the response amplitude.

\subsection*{Gain factor}

\begin{figure}
\begin{centering}
\includegraphics[width=1\columnwidth]{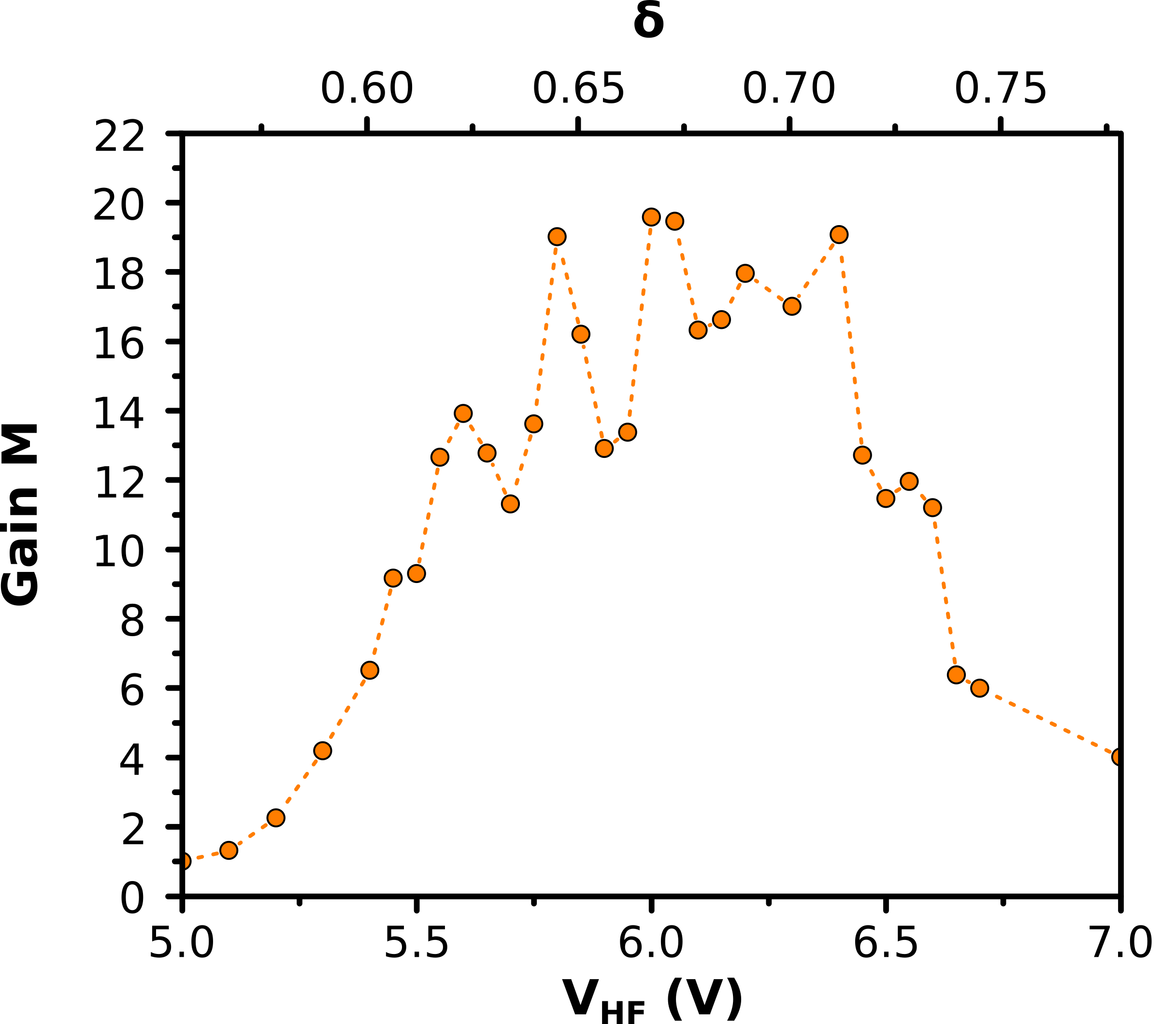}
\par\end{centering}
\caption{\label{Fig4}Gain factor $M$ as a function of the amplitude of the
external driving.}
\end{figure}

The gain or amplification factor can be inferred by quantifying the
achieved spectral power amplification. For every time traces recorded
on a time scale of $600s$, Discrete Fourier Transform (DFT) are performed.
The resulting DFT spectra are presented in Fig. \ref{Fig2}. They
feature peaks, the most prominent being at the modulation frequency
$\nu_{m}$. The achieved gain $M$ is then given by the ratio between
the strength of the peak in the DFT spectrum at $\nu_{m}$ for a given
amplitude of the external driving and its strength without external
driving ($V_{HF}=0\,V$). The induced gain factor is presented in
Figure \ref{Fig4}. The gain factor features a resonant-like behaviour:
The gain factor first rapidly rises with the strength of the external
driving, reaches a maximum for $V_{HF}=6.4\,V$ and then drops. The
maximum achieved gain factor is $M=20$.

Vibrational resonance is governed by an amplitude condition. It occurs
close to the transition from bistability to monostability, during
which the effective potential of the slow variable evolves from a
rapidly oscillating double well to a single well with a parametric
dependence on the high-frequency signal amplitude and frequency \citet{BaltanasPRE03}.
As such, this phenomenon has some features in common with parametric
amplification near the critical point. The shape of gain achieved
in vibrational resonance consequently reflects the (here symmetrical)
shape of the parametric gain in the system.

\subsection*{Theoretical analysis}

To figure out the origin of this resonant response, we introduce a
simplified theoretical model and compare its results to our experimental
findings. The original treatment of vibrational resonance in \citet{LandaJoPAMaG00,GittermanJoPAMaG01}
considers the motion of a nonlinear oscillator in a bistable potential,
subject to a low-frequency signal and a high-frequency drive. Theoretical
studies so far have mostly concentrated on studying the impact of
the potential shape on the resonance \citet{BaltanasPRE03,Roy-LayindeCAIJoNS16,VincentPRE18},
or the response to multi-frequency signals \citet{RajamaniCiNSaNS14}.
Interestingly, it was also noted in \citet{GittermanJoPAMaG01} that
one particularly important aspect of vibrational resonance was the
ability to change the stability of some equilibria, or to have control
over the shift of the resonance frequency. Our system is a nonlinear
oscillator with a small nonlinearity. It cannot show a bistable response
\emph{per se}, whatever the sign of the stiffness parameter $\alpha$.
However, with a quasi-resonant harmonic forcing, the nonlinear oscillator
can become bistable. It is then interesting to examine in more details
if an additional \textquotedbl high\textquotedbl{} frequency forcing
can induce a resonance on a small amplitude signal.

The nanoelectromechanical system can be described in a good approximation
as a forced nonlinear (cubic) Duffing oscillator \citet{ChowdhuryPRL17}.
Its dynamics can be modelled, in the limit of the small injection
and the dissipation of energy by

\begin{multline}
\ddot{x}+\eta\dot{x}+\omega_{o}^{2}x+\alpha x^{3}=\\
F\left[1+\gamma\cos\left(\omega_{m}t\right)+\delta\cos(\Omega t)\right]\cos\left(\omega_{\mathrm{f}}t\right),\label{Eq-Duffing-1}
\end{multline}
where $x(t)$ accounts for the out-of-plane displacement of the membrane,
$\mu$ is the effective damping, $\omega_{o}/2\pi$ is the natural
oscillation frequency of the membrane, $\alpha$ is the nonlinear
stiffness coefficient, $F$ is the amplitude of the modulated forcing
with frequency $\omega_{\mathrm{f}}/2\pi\equiv(\omega_{0}+\Delta)/2\pi$,
introducing the small detuning from resonance $\Delta$. The high
frequency amplitude modulation has an amplitude $F\delta$ and a frequency
$\Omega/2\pi=\nu_{HF}$. The oscillation amplitude of the oscillator
is the result of the beating of two frequencies: one fast at $\Omega$
and one slow at $\omega_{m}$. The parameters $\gamma$ and $\delta$
characterize the amplitude of the beating. By considering the following
separation of timescales for the forcing frequencies $\omega_{m}\ll\Omega\ll\omega_{0}$,
an amplitude equation for the time-averaged dynamics can be derived
(see Methods). We start by deriving the equation for the amplitude
of the forced nonlinear oscillator close to resonance ($\omega_{\mathrm{f}}\sim\omega_{0}$)
by looking for a solution in the form $x(t)=C(t)e^{i(\omega_{0}+\Delta)t}+cc$
(where $cc$ accounts for the complex conjugate term): 
\begin{multline}
\partial_{t}C=-\frac{\eta}{2}C-i\Delta C+i\frac{3\alpha}{2\omega_{0}}\vert C\vert^{2}C-\\
i\frac{F}{4\omega_{0}}\left(1+\gamma\cos\left(\omega_{m}t\right)+\delta\cos(\Omega t)\right),\label{eq:C}
\end{multline}
The strong timescale separation of the modulation frequencies motivates
the introduction of a time-averaged variable $A$ over the short period
$2\pi/\Omega$ \citet{Sanders07} such that 
\[
A(\tau)\equiv\frac{\Omega}{2\pi}\int_{\tau}^{\tau+2\pi/\Omega}C(t)dt.
\]
The amplitude equation for the averaged response writes

\begin{multline}
\partial_{\tau}A=-\frac{\eta}{2}A-i\left(\Delta-\frac{3\alpha F^{2}\delta^{2}}{16}\right)A+\\
i\frac{3\alpha}{2}\vert A\vert^{2}A-i\frac{F}{4}\left(1+\gamma\cos\left(\omega_{m}t\right)\right)\label{eq:Avg1}
\end{multline}
where we have introduced rescaled quantities: $\frac{F}{\omega_{0}}\to F$,
$\frac{\delta}{\Omega}\to\delta$ and $\frac{\alpha}{\omega_{0}}\to\alpha$.

The averaged equation satisfies an amplitude equation with a renormalised
detuning $\Delta-3\alpha F^{2}\delta^{2}/16$ which depends on the
high frequency driving amplitude. The most important aspect to note
here is that the non-resonant and \textquotedbl high\textquotedbl{}
frequency driving can modify the resonance behaviour of a nonlinear
system. To study how the intermediate frequency $\Omega$ modifies
the resonance region, we consider the polar representation $A=Re^{i\phi}/2$
with $\gamma=0$, and solve for the steady state $\dot{R}=\dot{\phi}=0$.
We get the characteristic equation

\begin{equation}
\frac{\eta^{2}}{4}R^{2}+\left[\left(\Delta-\frac{3\alpha F^{2}\delta^{2}}{16}\right)R-\frac{3}{8}\alpha R^{3}\right]^{2}=\frac{F^{2}}{4}\label{eq:amplVR}
\end{equation}
Note that in the limit of zero high-frequency amplitude modulation
($\delta\to0$) we recover the deterministic forced Duffing resonator
model. At this point, we highlight that the timescale separation hypothesis,
$\omega_{m}\ll\Omega\ll\omega_{0}$, is central to obtain this result.
Indeed, if we suppose $\omega_{m}\ll\omega_{0}\ll\Omega$, i.e. a
very high frequency driving and average Eq. (\ref{Eq-Duffing-1})
before deriving the amplitude equation, then we cannot show evidence
for vibrational resonance.

\begin{figure}
\begin{centering}
\includegraphics[width=1\columnwidth]{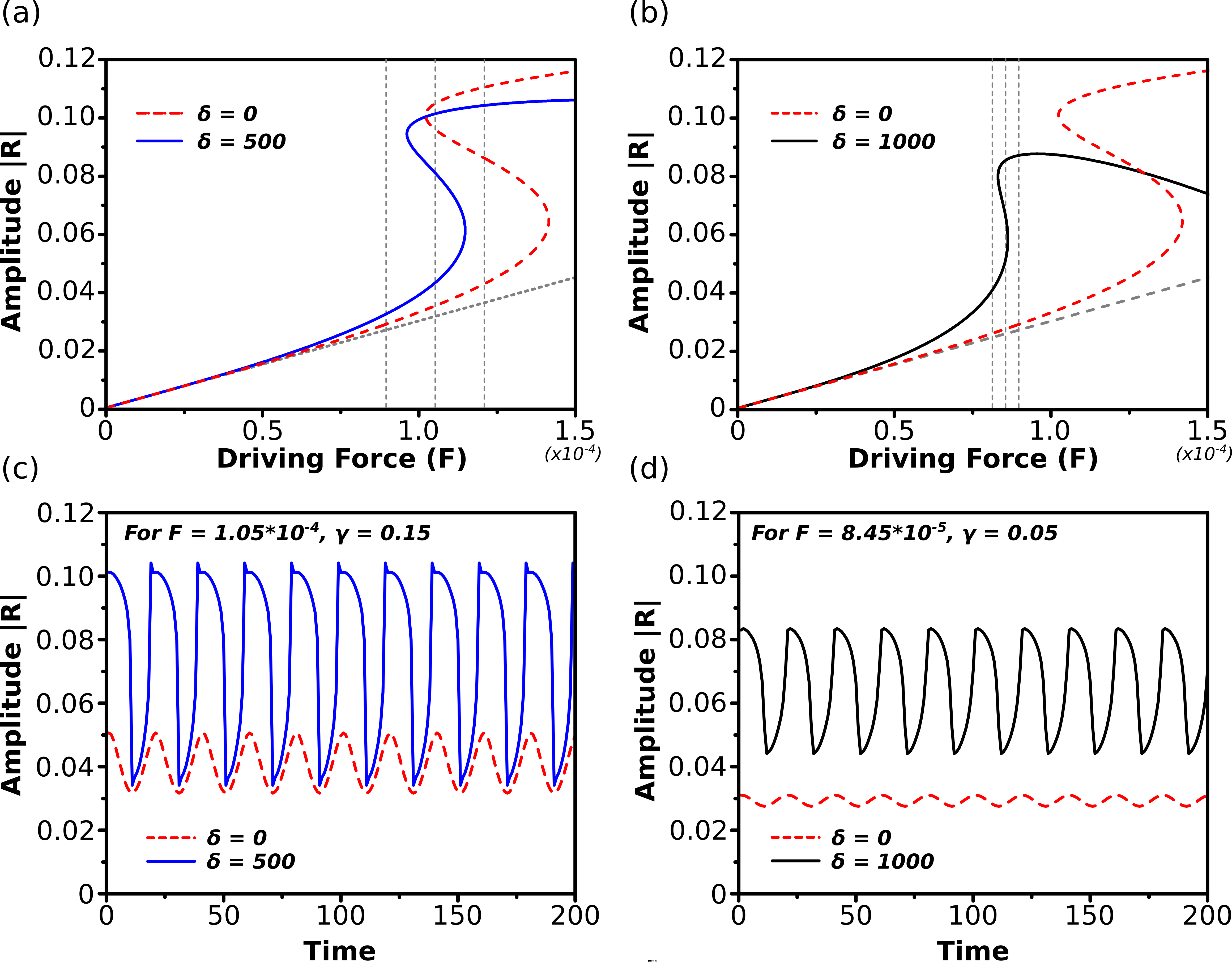}
\par\end{centering}
\caption{(a,b) : Steady state response $\vert R\vert$ versus F for (a) $\gamma=0.15$
and $\delta=0$ (red dashed lines) or $\delta=500$ (blue line), and
(b) $\gamma=0.05$ and $\delta=0$ (red dashed line) or $\delta=1000$
(black kine). Other parameters are $\sigma=0.0016$, $\eta=0.001$,
$\alpha=0.4$. The gray lines mark the linear regime (small driving
limit $F\to0$). (c,d) : time traces of the response $\vert R\vert$
corresponding to the maximum signal amplification in (a) and (b),
with respectively in (c) $F_{0}=1.05\times10^{-4}$ and $\delta=0$
(red dashed line) or $\delta=500$ (blue line) and in (d) $F_{0}=8.45\times10^{-5}$
and $\delta=0$ (red dashed line) or $\delta=1000$ (black line).
Vertical dashed lines in (a,b) lie at $F=F_{0}(1\pm\gamma)$ and $F=F_{0}$,
corresponding to the extreme and middle of the signal modulation.\label{fig:Steady-state-response}}
\end{figure}

We numerically investigate the amplitude equation given by Eq.(\ref{eq:amplVR})
for the parameters $\sigma=0.0016$, $\eta=0.001$, $\alpha=0.4$,
$\omega_{m}=2\pi/200000$. These parameters values are chosen to match
the experimental ones. By fitting the nonlinear resonance curve we
get the mechanical quality factor $Q_{M}=\eta^{-1}$ and the nonlinear
spring term $\alpha$ \citet{ChowdhuryPRL17}. The signal modulation
frequency is chosen to be much larger than mechanical quality factor
to ensure almost adiabatic evolution. Figure \ref{fig:Steady-state-response}
shows the steady-state response curves (inferred from Eq.(\ref{eq:amplVR}))
versus the driving amplitude for different high-frequency amplitude
modulation. Without any high frequency drive ($\delta=0$), the system
displays a large hysteretic response (see Fig.\ref{fig:Steady-state-response}a).
A slow modulation of amplitude less than the hysteresis width would
not produce any jump between the branches, hence would not produce
any strong amplification of the signal at $\omega_{m}$. The addition
of the high frequency drive introduces an extra detuning which deforms
the nonlinear response: in Fig.\ref{fig:Steady-state-response}a,
we observe that the center of the hysteresis loop is shifted towards
lower driving forces $F$, and that the width of the hysteresis shrinks
as well. Since the signal modulation amplitude scales as $F\gamma$,
this means that a smaller slow modulation amplitude $\gamma$ will
be necessary to overcome the hysteresis width and produce large jumps
between the lower and upper branch. This is the essence of the vibrational
resonance phenomenon. In order to check this, we integrated numerically
Eq.(\ref{eq:amplVR}) with a slow amplitude signal at $\omega_{m}$.
The results are shown on the time traces plotted in Figs.\ref{fig:Steady-state-response}{a,c}.
In the absence of high frequency amplitude modulation ($\delta=0$)
the response is quasi linear since the system cannot jump between
the lower and higher branches for the chosen modulation amplitude.
The amplification factor is close to one if the system resides on
the lower branch (same as the linear regime), or can be even smaller
if it resides in the upper branch (de-amplification). When the driving
$\delta$ is increased, the system is tuned into resonance and undergoes
synchronous jumps with the signal driving frequency between the lower
and upper branches (Fig.\ref{fig:Steady-state-response}c). This corresponds
to a large signal amplification, provided the signal amplitude is
large enough, i.e. larger than the hysteresis width for the chosen
parameters.

\begin{figure}
\begin{centering}
\includegraphics[width=1\columnwidth]{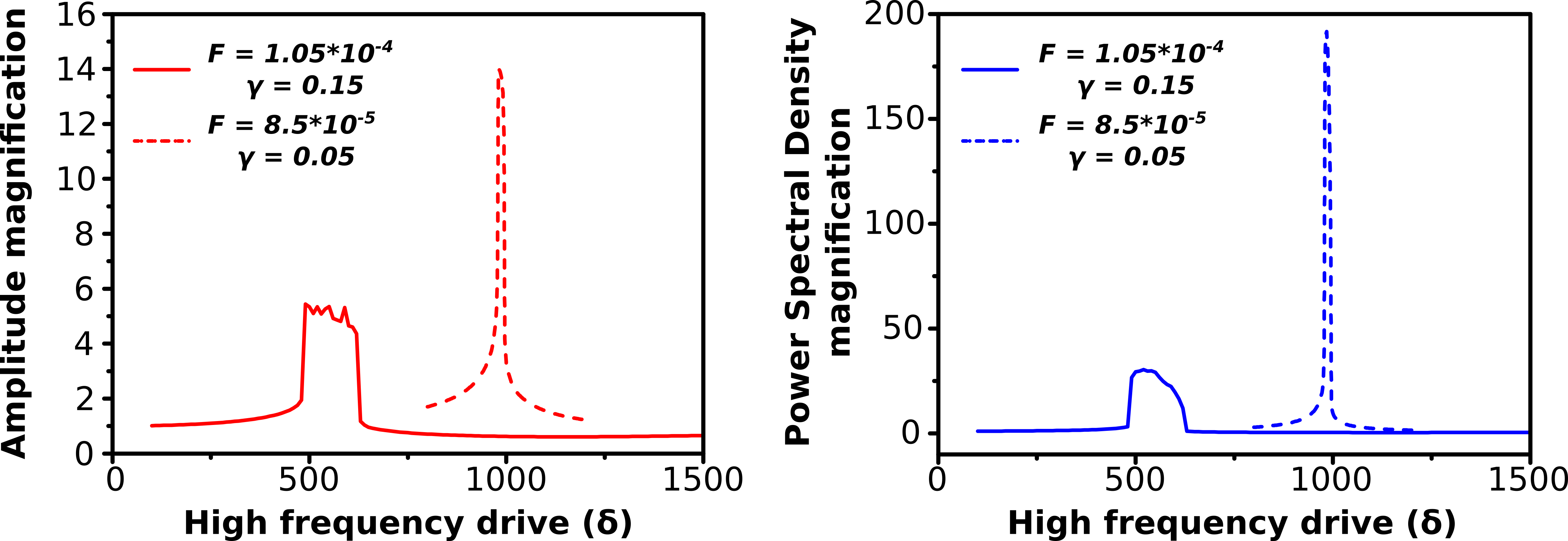}
\par\end{centering}
\caption{Left : Amplitude magnification as a function of the high frequency
drive. Right: Power spectral density amplification $M=PSD(\omega,\delta)/PSD(\omega,0)$
as a function of the high frequency drive. Parameters for the solid
red and blue lines are : $\sigma=0.0016$, $\eta=0.001$, $\alpha=0.4$,
$\omega=2\pi/200000$, $F=1.05\times10^{-4}$ and $\gamma=0.15$.
The parameters for the dashed red and blue lines are the same except
$F=8.5\times10^{-5}$ and $\gamma=0.05$ (lower resonant and low frequency
drivings). \label{fig:response}}
\end{figure}

The amplification is shown in Fig.\ref{fig:response}, rescaled to
the response with zero high-frequency drive in the quasi linear case,
i.e. on the lower branch. We plot both the amplitude ratio $M_{a}$
(ratio of the response amplitude with and without high frequency drive)
and the power-spectral density (PSD) ratio $M$. $M_{a}$ displays
a sharp transition corresponding to the tuning of the system into
the bistable region. When the high-frequency drive is not large enough,
the system stays in the lower branch and the response is quasi-linear,
leading to an amplification factor close to $1$. When the bistable
regime is reached, a large amplitude amplification is obtained. And
for still higher $\delta$ the system stays in the upper branch where
the response is sublinear, thus leading to de-amplification as expected.
The same thing occurs in the PSD, except that the transition is less
marked because the response of the system is highly nonlinear, hence
the spectral energy is spread among the different harmonics. Note
that we reach here, with the chosen parameters, a PSD amplification
of the same order of magnitude as in the experiment. However, much
larger amplification factors can be reached for other slightly different
parameters, as illustrated in Fig. \ref{fig:response} where $M\sim140$
is obtained for still smaller linear driving signal not accessible
in experiments. This important point is illustrated in Figs.\ref{fig:Steady-state-response}b,d.
If the signal strength is too small to overcome the hysteresis width,
it is possible to increase the high frequency drive to tune it into
the resonance. As shown in Fig.\ref{fig:Steady-state-response}b for
$\delta=1000$, a higher high-frequency modulation shifts the hysteresis
curve further to the right, i.e. to lower overall forcing, but most
importantly reduces the width of the hysteresis while not changing
the hysteresis height too much. This makes it possible to amplify
a much weaker signal by the vibrational resonance phenomenon. Note
also that the amplification factor is even much larger in that case
because of the already discussed different effect on the width and
on the height of the hysteresis. This shows that it is necessary to
tune both the high frequency drive $\delta$ and the modulation strength
$F$ to amplify optimally a signal of a given amplitude.

\section*{Discussion}

The previous analyses clearly indicate the primordial role of the
high frequency amplitude modulation and of the proper timescale separation
in such vibrational resonance phenomenon. The former allows to control
the nonlinear resonance in order to amplify weak signals. The latter,
while being compulsory for technical reasons in the theoretical analysis,
could potentially be relaxed in experiments. The exact value of the
external drive frequency $\Omega$ is not critical at all, as long
as it satisfies the timescale separation condition with $\omega_{m}\ll\omega_{\mathrm{f}},\Omega$
and if it remains non-resonant. Signal amplification results from
the tuning into resonance of the non-linear response of the system.
Here we have shown its effect on a bistable response curve, but amplification
could occur also in the case of a non multivalued response, as long
as the slope of the tuned response is large enough to ensure appropriate
amplification. In principle, for any given signal amplitude, it is
always possible to adjust both the forcing strength $F$ and the high
frequency modulation strength $\delta$ for vibrational resonance
to occur. However, the maximum gain achievable will be a complicated
function of all the system parameters.

By comparing the amplitude magnification curves in Figs.\ref{Fig4}
and \ref{fig:response} we note a slight softening effect on the experimental
gain curve\, whereas the theoretical one shows a sharp transition
to high gain when the signal modulation is larger than the hysteresis
curve turning points. This difference can be attributed to residual
noise in the experiment which can modify the behaviour of the system
close to the turning points of the nonlinear response.

At last, we can compare also the results of vibrational resonance
with respect to those of stochastic resonance, even though the physics
involved is quite different. As observed in previous optical implementations
of vibrational resonance \citet{ChizhevskyIJBC08,ChizhevskyPRA05}
in non-parametrically forced bistable systems, the time series obtained
in the vibrational resonance scenario suggest a higher gain factor
with respect to the one observed for our system in stochastic resonance
\citet{ChowdhuryPRL17}, hence supporting the potential of the technique
of vibrational resonance for small signal amplification.

In conclusion, we established and analysed the conditions for using
vibrational resonance in order to enhance weak signals in a forced
nonlinear oscillator, even if the system is initially monostable.
The physical phenomenon is based on the resonance manipulation thanks
to a non-resonant, high frequency amplitude-modulation drive obeying
a timescale separation condition.We derived a model to describe vibrational
resonance in a monostable, forced nonlinear oscillator which shows
good agreement with our experimental results obtained on a forced
nano-electromechanical membrane. This deterministic amplification
method gives rise to high amplification factors, especially when compared
to stochastic resonance \citet{RenPRE17}. In a more general framework,
it may open new avenues for the manipulation of non-linear resonances
with the addition of a non-resonant driving field .

\section*{Aknowledgements}

This work is supported by the French RENATECH network, the Marie Curie
Innovative Training Networks (ITN) cQOM\textcolor{black}{{} and the
European Union's Horizon 2020 research and innovation program under
grant agreement No 732894 (FET Proactive HOT)}. M.G.C. thanks the
Millennium Institute for Research in Optics (MIRO) and FONDECYT projects
Grants No. 1180903 for financial support.

\section*{Methods}

\paragraph*{Fabrication of the InP resonator membrane}

The fabrication of the whole platform is based on a 3D heterogeneous
integration process involving mainly four steps. First, a 400 nm-thick
SiO$_{2}$ layer is deposited on the 260 nm-thick InP membrane, which
is grown along with a 500 nm-thick InGaAs etch-stop layer on top of
an InP (100) substrate by metal organic vapour phase epitaxy. Simultaneously,
interdigitated electrodes (IDTs) arrays, displaying a finger period
of 2 $\mu m$, finger length of 10 $\mu m$ and electrode width of
500 nm, are deposited on a Si substrate. The patterning process involves
an electronic lithography, deposition of a 200 nm-thick gold layer
and standard lift-off. The Si chip is then spin-coated with a 200
nm-thick DiVinylSiloxane-BenzoCycloButene (DVS-BCB) layer, thereby
planarising the Si substrate. In the second step, the InP wafer is
bonded on the Si substrate at high-temperature (300$^{\circ}\mathrm{C}$)
by positioning the SiO$_{2}$ layer atop the DVS-BCB layer and by
using a vacuum wafer bonding technique \citet{KarleAPL10}. The InP
substrate and InGaAs etch-stop layer are then removed by chemical
etching, leaving the residual 260 nm-thick InP membrane on the DVS-BCB-SiO$_{2}$
layer. In the third step, the InP membrane is patterned by standard
e-beam lithography and dry-etching, to form a two-dimensional square-lattice
photonic crystal of periodicity 532 nm, hole radius 181 nm and whole
surface of 10$\times$20 $\mu m^{2}$. It is clamped by four tethers
of 2 $\mu m$ length and 1 $\mu m$ width, in order to reduce clamping
losses. The alignment of the photonic crystal mirror with respect
to the IDT's arrays, is performed with an accuracy better than 20
nm, by making use of alignment marks deposited beforehand on the Si
substrate. Last, the photonic crystal membranes are released by under-etching
the underlying 400 nm-thick SiO$_{2}$ layer, followed by a critical
point drying step. The lateral InP suspension pads act as protective
structures for the SiO$_{2}$ layer beneath them, leaving them anchored
to the substrate.

\paragraph*{Measurement of the out-of-plane motion}

A He-Ne laser with wavelength of $633\,nm$ is sent to the membrane.
The reflectivity of the membrane is enhanced up to $50\%$ by piercing
a square lattice photonic crystal in it \citet{MaklesOL15}. The laser
is focused on the membrane via an objective with a NA of $0.4$. The
light reflected by the membrane is brought to interference with a
strong local oscillator. A balanced homodyne detector locked on the
drive frequency at the interferometer output is then used to decipher
the amplitude and phase of the mechanical motion.

\paragraph*{Actuation of the mechanical oscillator}

The membrane is driven via the electrostatic force induced by the
electrodes placed underneath. These electrodes are connected to an
external signal generator which can go up to 50 MHz and is synchronised
with a lock-in amplifier (HF2LI) which demodulates the detected signal
at the actuation frequency. For vibrational resonance, the weak ($\nu_{m}$)
modulated signal is generated by an another signal generator (Model
Agilent 33522A) and combined with the additive ($\nu_{HF}$) signal
in the HF2LI. This signal is then modulated at the frequency of the
quasi-resonant forcing ($\nu_{\mathrm{f}}$) and sent to the electrodes.
The electrical signal converted by the photodiodes (Thorlabs APD120A2)
is time-recorded with the oscilloscope function of the HF2LI.

\paragraph*{Derivation of the theoretical model}

Let us consider the timescale separation $\omega_{m}\ll\Omega\ll\omega_{0}$
and the resonance condition $\omega_{\mathrm{f}}\sim\omega_{0}$.
We first look for a solution to Eq.~(\ref{Eq-Duffing-1}) using the
ansatz $x(t)=C(t)e^{i(\omega_{0}+\Delta)t}+cc$ (where $cc$ accounts
for the complex conjugate term). After straightforward algebra, one
gets an amplitude equation for the slow envelope $C(t)$, assuming
that $\partial_{tt}C\ll\omega_{0}^{2}C$ and $\partial_{t}C\ll\omega_{0}C$:
\begin{multline}
\partial_{t}C=-\frac{\eta}{2}C-i\Delta C+i\frac{3\alpha}{2\omega_{0}}\vert C\vert^{2}C-\\
i\frac{F}{4\omega_{0}}\left(1+\gamma\cos\left(\omega_{m}t\right)+\delta\cos(\Omega t)\right),\label{eq:C-1}
\end{multline}
The amplitude equation (Eq.(\ref{eq:C-1})) corresponds to the one
of a forced oscillator with temporally modulated amplitude. Since
we have a strong timescale separation of the modulation frequencies,
$\omega_{m}\ll\Omega$, one can consider the averaged variable on
the short period $2\pi/\Omega$ 
\[
A(\tau)\equiv\frac{\Omega}{2\pi}\int_{\tau}^{\tau+2\pi/\Omega}C(t)dt
\]
and homogenise the scales \citet{Bogoliubov61} by writing 
\begin{equation}
C(t)=A(\tau)-\frac{\delta F}{4\omega_{0}\Omega}e^{i\Omega t}\label{eq:Homogenization-1}
\end{equation}
Considering that the envelope $A(\tau)$ is a slow variable ($\partial_{\tau}A\ll\omega_{m}A$),
using the ansatz (\ref{eq:Homogenization-1}) in equation (\ref{eq:C-1})
and averaging over the period $2\pi/\Omega$, we get the amplitude
equation for the averaged response 
\begin{multline}
\partial_{\tau}A=-\frac{\eta}{2}A-i\left(\Delta-\frac{3\alpha F^{2}\delta^{2}}{16\omega_{0}^{3}\Omega^{2}}\right)A+\\
i\frac{3\alpha}{2\omega_{0}}\vert A\vert^{2}A-i\frac{F}{4\omega_{0}}\left(1+\gamma\cos\left(\omega_{m}t\right)\right)\label{eq:Avg-1}
\end{multline}
Introducing the notation $F'=\frac{F}{\omega_{0}}$, $\delta'=\frac{\delta}{\Omega}$
and $\alpha'=\frac{\alpha}{\omega_{0}}$ and omitting $'$, the equation
reads

\begin{multline}
\partial_{\tau}A=-\frac{\eta}{2}A-i\left(\Delta-\frac{3\alpha F^{2}\delta^{2}}{16}\right)A+\\
i\frac{3\alpha}{2}\vert A\vert^{2}A-i\frac{F}{4}\left(1+\gamma\cos\left(\omega_{m}t\right)\right)\label{eq:Avg1-1}
\end{multline}
We further introduce a Madelung transform $A=Re^{i\phi}/2$ and $\gamma=0$,
and get

\begin{eqnarray}
\dot{R}=-\frac{\eta}{2}R-\frac{F}{2}\sin(\phi)\label{eq:array-1}\\
R\dot{\phi}=-\left(\Delta-\frac{3\alpha F^{2}\delta^{2}}{16}\right)R+\frac{3}{8}\alpha R^{3}-\frac{F}{2}\cos(\phi)
\end{eqnarray}

At steady state,$\dot{R}=\dot{\phi}=0$ and we finally get the characteristic
equation

\begin{equation}
\frac{\eta^{2}}{4}R^{2}+\left[\left(\Delta-\frac{3\alpha F^{2}\delta^{2}}{16}\right)R-\frac{3}{8}\alpha R^{3}\right]^{2}=\frac{F^{2}}{4}\label{eq:amplVR-1}
\end{equation}

\section*{References}

\bibliographystyle{apsrev4-1}
\bibliography{5C__Users_braive-adm_Documents_OptoMechanics_Article_Vibrational_Res_Version5_bib-VR}

\begin{thebibliography}{28}%
\makeatletter
\providecommand \@ifxundefined [1]{%
 \@ifx{#1\undefined}
}%
\providecommand \@ifnum [1]{%
 \ifnum #1\expandafter \@firstoftwo
 \else \expandafter \@secondoftwo
 \fi
}%
\providecommand \@ifx [1]{%
 \ifx #1\expandafter \@firstoftwo
 \else \expandafter \@secondoftwo
 \fi
}%
\providecommand \natexlab [1]{#1}%
\providecommand \enquote  [1]{``#1''}%
\providecommand \bibnamefont  [1]{#1}%
\providecommand \bibfnamefont [1]{#1}%
\providecommand \citenamefont [1]{#1}%
\providecommand \href@noop [0]{\@secondoftwo}%
\providecommand \href [0]{\begingroup \@sanitize@url \@href}%
\providecommand \@href[1]{\@@startlink{#1}\@@href}%
\providecommand \@@href[1]{\endgroup#1\@@endlink}%
\providecommand \@sanitize@url [0]{\catcode `\\12\catcode `\$12\catcode
  `\&12\catcode `\#12\catcode `\^12\catcode `\_12\catcode `\%12\relax}%
\providecommand \@@startlink[1]{}%
\providecommand \@@endlink[0]{}%
\providecommand \url  [0]{\begingroup\@sanitize@url \@url }%
\providecommand \@url [1]{\endgroup\@href {#1}{\urlprefix }}%
\providecommand \urlprefix  [0]{URL }%
\providecommand \Eprint [0]{\href }%
\providecommand \doibase [0]{http://dx.doi.org/}%
\providecommand \selectlanguage [0]{\@gobble}%
\providecommand \bibinfo  [0]{\@secondoftwo}%
\providecommand \bibfield  [0]{\@secondoftwo}%
\providecommand \translation [1]{[#1]}%
\providecommand \BibitemOpen [0]{}%
\providecommand \bibitemStop [0]{}%
\providecommand \bibitemNoStop [0]{.\EOS\space}%
\providecommand \EOS [0]{\spacefactor3000\relax}%
\providecommand \BibitemShut  [1]{\csname bibitem#1\endcsname}%
\let\auto@bib@innerbib\@empty
\bibitem [{\citenamefont {Benzi}\ \emph {et~al.}(1981)\citenamefont {Benzi},
  \citenamefont {Sutera},\ and\ \citenamefont {Vulpiani}}]{BenziJPAMG81}%
  \BibitemOpen
  \bibfield  {author} {\bibinfo {author} {\bibfnamefont {R.}~\bibnamefont
  {Benzi}}, \bibinfo {author} {\bibfnamefont {A.}~\bibnamefont {Sutera}}, \
  and\ \bibinfo {author} {\bibfnamefont {A.}~\bibnamefont {Vulpiani}},\
  }\href@noop {} {\bibfield  {journal} {\bibinfo  {journal} {J. Phys. A: Math.
  Gen,}\ }\textbf {\bibinfo {volume} {14}},\ \bibinfo {pages} {453} (\bibinfo
  {year} {1981})}\BibitemShut {NoStop}%
\bibitem [{\citenamefont {Landa}\ and\ \citenamefont
  {McClintock}(2000)}]{LandaJoPAMaG00}%
  \BibitemOpen
  \bibfield  {author} {\bibinfo {author} {\bibfnamefont {P.~S.}\ \bibnamefont
  {Landa}}\ and\ \bibinfo {author} {\bibfnamefont {P.~V.~E.}\ \bibnamefont
  {McClintock}},\ }\href {\doibase 10.1088/0305-4470/33/45/103} {\bibfield
  {journal} {\bibinfo  {journal} {J. Phys. A: Math. Gen.}\ }\textbf {\bibinfo
  {volume} {33}},\ \bibinfo {pages} {L433} (\bibinfo {year}
  {2000})}\BibitemShut {NoStop}%
\bibitem [{\citenamefont {Zaikin}\ \emph {et~al.}(2002)\citenamefont {Zaikin},
  \citenamefont {L\'opez}, \citenamefont {Baltan\'as}, \citenamefont {Kurths},\
  and\ \citenamefont {Sanju\'an}}]{ZaikinPRE02}%
  \BibitemOpen
  \bibfield  {author} {\bibinfo {author} {\bibfnamefont {A.~A.}\ \bibnamefont
  {Zaikin}}, \bibinfo {author} {\bibfnamefont {L.}~\bibnamefont {L\'opez}},
  \bibinfo {author} {\bibfnamefont {J.~P.}\ \bibnamefont {Baltan\'as}},
  \bibinfo {author} {\bibfnamefont {J.}~\bibnamefont {Kurths}}, \ and\ \bibinfo
  {author} {\bibfnamefont {M.~A.~F.}\ \bibnamefont {Sanju\'an}},\ }\href
  {\doibase 10.1103/PhysRevE.66.011106} {\bibfield  {journal} {\bibinfo
  {journal} {Phys. Rev. E}\ }\textbf {\bibinfo {volume} {66}},\ \bibinfo
  {pages} {011106} (\bibinfo {year} {2002})}\BibitemShut {NoStop}%
\bibitem [{\citenamefont {Fauve}\ and\ \citenamefont
  {Heslot}(1983)}]{FauvePLA83}%
  \BibitemOpen
  \bibfield  {author} {\bibinfo {author} {\bibfnamefont {S.}~\bibnamefont
  {Fauve}}\ and\ \bibinfo {author} {\bibfnamefont {F.}~\bibnamefont {Heslot}},\
  }\href {\doibase 10.1016/0375-9601(83)90086-5} {\bibfield  {journal}
  {\bibinfo  {journal} {Phys. Lett. A}\ }\textbf {\bibinfo {volume} {97}},\
  \bibinfo {pages} {5 } (\bibinfo {year} {1983})}\BibitemShut {NoStop}%
\bibitem [{\citenamefont {McNamara}\ \emph {et~al.}(1988)\citenamefont
  {McNamara}, \citenamefont {Wiesenfeld},\ and\ \citenamefont
  {Roy}}]{McNamaraPRL88}%
  \BibitemOpen
  \bibfield  {author} {\bibinfo {author} {\bibfnamefont {B.}~\bibnamefont
  {McNamara}}, \bibinfo {author} {\bibfnamefont {K.}~\bibnamefont
  {Wiesenfeld}}, \ and\ \bibinfo {author} {\bibfnamefont {R.}~\bibnamefont
  {Roy}},\ }\href@noop {} {\bibfield  {journal} {\bibinfo  {journal} {Phys.
  Rev. Lett.}\ }\textbf {\bibinfo {volume} {60}},\ \bibinfo {pages} {2626}
  (\bibinfo {year} {1988})}\BibitemShut {NoStop}%
\bibitem [{\citenamefont {Barbay}\ \emph {et~al.}(2000)\citenamefont {Barbay},
  \citenamefont {Giacomelli},\ and\ \citenamefont {Marin}}]{BarbayPRE00}%
  \BibitemOpen
  \bibfield  {author} {\bibinfo {author} {\bibfnamefont {S.}~\bibnamefont
  {Barbay}}, \bibinfo {author} {\bibfnamefont {G.}~\bibnamefont {Giacomelli}},
  \ and\ \bibinfo {author} {\bibfnamefont {F.}~\bibnamefont {Marin}},\ }\href
  {\doibase 10.1103/PhysRevE.61.157} {\bibfield  {journal} {\bibinfo  {journal}
  {Phys. Rev. E}\ }\textbf {\bibinfo {volume} {61}},\ \bibinfo {pages} {157}
  (\bibinfo {year} {2000})}\BibitemShut {NoStop}%
\bibitem [{\citenamefont {Chizhevsky}\ \emph {et~al.}(2003)\citenamefont
  {Chizhevsky}, \citenamefont {Smeu},\ and\ \citenamefont
  {Giacomelli}}]{ChizhevskyPRL03}%
  \BibitemOpen
  \bibfield  {author} {\bibinfo {author} {\bibfnamefont {V.~N.}\ \bibnamefont
  {Chizhevsky}}, \bibinfo {author} {\bibfnamefont {E.}~\bibnamefont {Smeu}}, \
  and\ \bibinfo {author} {\bibfnamefont {G.}~\bibnamefont {Giacomelli}},\
  }\href {\doibase 10.1103/PhysRevLett.91.220602} {\bibfield  {journal}
  {\bibinfo  {journal} {Phys. Rev. Lett.}\ }\textbf {\bibinfo {volume} {91}},\
  \bibinfo {pages} {220602} (\bibinfo {year} {2003})}\BibitemShut {NoStop}%
\bibitem [{\citenamefont {Chizhevsky}\ and\ \citenamefont
  {Giacomelli}(2005)}]{ChizhevskyPRA05}%
  \BibitemOpen
  \bibfield  {author} {\bibinfo {author} {\bibfnamefont {V.~N.}\ \bibnamefont
  {Chizhevsky}}\ and\ \bibinfo {author} {\bibfnamefont {G.}~\bibnamefont
  {Giacomelli}},\ }\href {\doibase 10.1103/PhysRevA.71.011801} {\bibfield
  {journal} {\bibinfo  {journal} {Phys. Rev. A}\ }\textbf {\bibinfo {volume}
  {71}},\ \bibinfo {pages} {011801} (\bibinfo {year} {2005})}\BibitemShut
  {NoStop}%
\bibitem [{\citenamefont {Ullner}\ \emph {et~al.}(2003)\citenamefont {Ullner},
  \citenamefont {Zaikin}, \citenamefont {Garc{\i}\'a-Ojalvo}, \citenamefont
  {B\'ascones},\ and\ \citenamefont {Kurths}}]{UllnerPLA03}%
  \BibitemOpen
  \bibfield  {author} {\bibinfo {author} {\bibfnamefont {E.}~\bibnamefont
  {Ullner}}, \bibinfo {author} {\bibfnamefont {A.}~\bibnamefont {Zaikin}},
  \bibinfo {author} {\bibfnamefont {J.}~\bibnamefont {Garc{\i}\'a-Ojalvo}},
  \bibinfo {author} {\bibfnamefont {R.}~\bibnamefont {B\'ascones}}, \ and\
  \bibinfo {author} {\bibfnamefont {J.}~\bibnamefont {Kurths}},\ }\href
  {\doibase 10.1016/s0375-9601(03)00681-9} {\bibfield  {journal} {\bibinfo
  {journal} {Phys. Lett. A}\ }\textbf {\bibinfo {volume} {312}},\ \bibinfo
  {pages} {348} (\bibinfo {year} {2003})}\BibitemShut {NoStop}%
\bibitem [{\citenamefont {Douglass}\ \emph {et~al.}(1993)\citenamefont
  {Douglass}, \citenamefont {Wilkens}, \citenamefont {Pantazelou},\ and\
  \citenamefont {Moss}}]{DouglassN93}%
  \BibitemOpen
  \bibfield  {author} {\bibinfo {author} {\bibfnamefont {J.~K.}\ \bibnamefont
  {Douglass}}, \bibinfo {author} {\bibfnamefont {L.}~\bibnamefont {Wilkens}},
  \bibinfo {author} {\bibfnamefont {E.}~\bibnamefont {Pantazelou}}, \ and\
  \bibinfo {author} {\bibfnamefont {F.}~\bibnamefont {Moss}},\ }\href@noop {}
  {\bibfield  {journal} {\bibinfo  {journal} {Nature}\ }\textbf {\bibinfo
  {volume} {365}},\ \bibinfo {pages} {337} (\bibinfo {year}
  {1993})}\BibitemShut {NoStop}%
\bibitem [{\citenamefont {Badzey}\ and\ \citenamefont
  {Mohanty}(2005)}]{BadzeyNat05}%
  \BibitemOpen
  \bibfield  {author} {\bibinfo {author} {\bibfnamefont {R.~L.}\ \bibnamefont
  {Badzey}}\ and\ \bibinfo {author} {\bibfnamefont {P.}~\bibnamefont
  {Mohanty}},\ }\href@noop {} {\bibfield  {journal} {\bibinfo  {journal}
  {Nature}\ }\textbf {\bibinfo {volume} {437}},\ \bibinfo {pages} {995}
  (\bibinfo {year} {2005})}\BibitemShut {NoStop}%
\bibitem [{\citenamefont {Guerra}\ \emph {et~al.}(2009)\citenamefont {Guerra},
  \citenamefont {Dunn},\ and\ \citenamefont {Mohanty}}]{GuerraNL09}%
  \BibitemOpen
  \bibfield  {author} {\bibinfo {author} {\bibfnamefont {D.~N.}\ \bibnamefont
  {Guerra}}, \bibinfo {author} {\bibfnamefont {T.}~\bibnamefont {Dunn}}, \ and\
  \bibinfo {author} {\bibfnamefont {P.}~\bibnamefont {Mohanty}},\ }\href
  {\doibase 10.1021/nl9004546} {\bibfield  {journal} {\bibinfo  {journal} {Nano
  Lett.}\ }\textbf {\bibinfo {volume} {9}},\ \bibinfo {pages} {3096} (\bibinfo
  {year} {2009})}\BibitemShut {NoStop}%
\bibitem [{\citenamefont {Venstra}\ \emph {et~al.}(2013)\citenamefont
  {Venstra}, \citenamefont {Westra},\ and\ \citenamefont {van~der
  Zant}}]{VenstraNatCom13}%
  \BibitemOpen
  \bibfield  {author} {\bibinfo {author} {\bibfnamefont {W.~J.}\ \bibnamefont
  {Venstra}}, \bibinfo {author} {\bibfnamefont {H.~J.~R.}\ \bibnamefont
  {Westra}}, \ and\ \bibinfo {author} {\bibfnamefont {H.~S.~J.}\ \bibnamefont
  {van~der Zant}},\ }\href {http://dx.doi.org/10.1038/ncomms3624} {\bibfield
  {journal} {\bibinfo  {journal} {Nat. Commun.}\ }\textbf {\bibinfo {volume}
  {4}},\  (\bibinfo {year} {2013})}\BibitemShut {NoStop}%
\bibitem [{\citenamefont {Chowdhury}\ \emph {et~al.}(2017)\citenamefont
  {Chowdhury}, \citenamefont {Barbay}, \citenamefont {Clerc}, \citenamefont
  {Robert-Philip},\ and\ \citenamefont {Braive}}]{ChowdhuryPRL17}%
  \BibitemOpen
  \bibfield  {author} {\bibinfo {author} {\bibfnamefont {A.}~\bibnamefont
  {Chowdhury}}, \bibinfo {author} {\bibfnamefont {S.}~\bibnamefont {Barbay}},
  \bibinfo {author} {\bibfnamefont {M.~G.}\ \bibnamefont {Clerc}}, \bibinfo
  {author} {\bibfnamefont {I.}~\bibnamefont {Robert-Philip}}, \ and\ \bibinfo
  {author} {\bibfnamefont {R.}~\bibnamefont {Braive}},\ }\href {\doibase
  10.1103/physrevlett.119.234101} {\bibfield  {journal} {\bibinfo  {journal}
  {Phys. Rev. Lett.}\ }\textbf {\bibinfo {volume} {119}} (\bibinfo {year}
  {2017}),\ 10.1103/physrevlett.119.234101}\BibitemShut {NoStop}%
\bibitem [{\citenamefont {Machura}\ \emph {et~al.}(2010)\citenamefont
  {Machura}, \citenamefont {Kostur},\ and\ \citenamefont
  {\L{}uczka}}]{MACHURA2010445}%
  \BibitemOpen
  \bibfield  {author} {\bibinfo {author} {\bibfnamefont {L.}~\bibnamefont
  {Machura}}, \bibinfo {author} {\bibfnamefont {M.}~\bibnamefont {Kostur}}, \
  and\ \bibinfo {author} {\bibfnamefont {J.}~\bibnamefont {\L{}uczka}},\ }\href
  {\doibase https://doi.org/10.1016/j.chemphys.2010.03.008} {\bibfield
  {journal} {\bibinfo  {journal} {Chemical Physics}\ }\textbf {\bibinfo
  {volume} {375}},\ \bibinfo {pages} {445 } (\bibinfo {year} {2010})},\
  \bibinfo {note} {stochastic processes in Physics and Chemistry (in honor of
  Peter H\"anggi)}\BibitemShut {NoStop}%
\bibitem [{\citenamefont {Kaplan}(2005)}]{Kaplan05}%
  \BibitemOpen
  \bibfield  {author} {\bibinfo {author} {\bibfnamefont {E.~D.}\ \bibnamefont
  {Kaplan}},\ }\href {\doibase 10.1017/S0373463300023730} {\emph {\bibinfo
  {title} {Understanding GPS: Principles and Applications}}}\ (\bibinfo
  {publisher} {Artech House},\ \bibinfo {year} {2005})\BibitemShut {NoStop}%
\bibitem [{\citenamefont {Chowdhury}\ \emph {et~al.}(2016)\citenamefont
  {Chowdhury}, \citenamefont {I.~Yeo}, \citenamefont {Raineri}, \citenamefont
  {Beaudoin}, \citenamefont {Sagnes}, \citenamefont {Raj}, \citenamefont
  {Robert-Philip},\ and\ \citenamefont {Braive}}]{ChowdhuryAPL16}%
  \BibitemOpen
  \bibfield  {author} {\bibinfo {author} {\bibfnamefont {A.}~\bibnamefont
  {Chowdhury}}, \bibinfo {author} {\bibfnamefont {V.~T.}\ \bibnamefont
  {I.~Yeo}}, \bibinfo {author} {\bibfnamefont {F.}~\bibnamefont {Raineri}},
  \bibinfo {author} {\bibfnamefont {G.}~\bibnamefont {Beaudoin}}, \bibinfo
  {author} {\bibfnamefont {I.}~\bibnamefont {Sagnes}}, \bibinfo {author}
  {\bibfnamefont {R.}~\bibnamefont {Raj}}, \bibinfo {author} {\bibfnamefont
  {I.}~\bibnamefont {Robert-Philip}}, \ and\ \bibinfo {author} {\bibfnamefont
  {R.}~\bibnamefont {Braive}},\ }\href@noop {} {\bibfield  {journal} {\bibinfo
  {journal} {Appl. Phys. Lett.}\ }\textbf {\bibinfo {volume} {108}},\ \bibinfo
  {pages} {163102} (\bibinfo {year} {2016})}\BibitemShut {NoStop}%
\bibitem [{\citenamefont {Baltan{\'{a}}s}\ \emph {et~al.}(2003)\citenamefont
  {Baltan{\'{a}}s}, \citenamefont {L{\'{o}}pez}, \citenamefont {Blechman},
  \citenamefont {Landa}, \citenamefont {Zaikin}, \citenamefont {Kurths},\ and\
  \citenamefont {Sanju{\'{a}}n}}]{BaltanasPRE03}%
  \BibitemOpen
  \bibfield  {author} {\bibinfo {author} {\bibfnamefont {J.~P.}\ \bibnamefont
  {Baltan{\'{a}}s}}, \bibinfo {author} {\bibfnamefont {L.}~\bibnamefont
  {L{\'{o}}pez}}, \bibinfo {author} {\bibfnamefont {I.~I.}\ \bibnamefont
  {Blechman}}, \bibinfo {author} {\bibfnamefont {P.~S.}\ \bibnamefont {Landa}},
  \bibinfo {author} {\bibfnamefont {A.}~\bibnamefont {Zaikin}}, \bibinfo
  {author} {\bibfnamefont {J.}~\bibnamefont {Kurths}}, \ and\ \bibinfo {author}
  {\bibfnamefont {M.~A.~F.}\ \bibnamefont {Sanju{\'{a}}n}},\ }\href {\doibase
  10.1103/physreve.67.066119} {\bibfield  {journal} {\bibinfo  {journal} {Phys.
  Rev. E}\ }\textbf {\bibinfo {volume} {67}} (\bibinfo {year} {2003}),\
  10.1103/physreve.67.066119}\BibitemShut {NoStop}%
\bibitem [{\citenamefont {Gitterman}(2001)}]{GittermanJoPAMaG01}%
  \BibitemOpen
  \bibfield  {author} {\bibinfo {author} {\bibfnamefont {M.}~\bibnamefont
  {Gitterman}},\ }\href {\doibase 10.1088/0305-4470/34/24/101} {\bibfield
  {journal} {\bibinfo  {journal} {J. Phys. A: Math. Gen.}\ }\textbf {\bibinfo
  {volume} {34}},\ \bibinfo {pages} {L355} (\bibinfo {year}
  {2001})}\BibitemShut {NoStop}%
\bibitem [{\citenamefont {Roy-Layinde}\ \emph {et~al.}(2016)\citenamefont
  {Roy-Layinde}, \citenamefont {Laoye}, \citenamefont {Popoola},\ and\
  \citenamefont {Vincent}}]{Roy-LayindeCAIJoNS16}%
  \BibitemOpen
  \bibfield  {author} {\bibinfo {author} {\bibfnamefont {T.~O.}\ \bibnamefont
  {Roy-Layinde}}, \bibinfo {author} {\bibfnamefont {J.~A.}\ \bibnamefont
  {Laoye}}, \bibinfo {author} {\bibfnamefont {O.~O.}\ \bibnamefont {Popoola}},
  \ and\ \bibinfo {author} {\bibfnamefont {U.~E.}\ \bibnamefont {Vincent}},\
  }\href {\doibase 10.1063/1.4962403} {\bibfield  {journal} {\bibinfo
  {journal} {Chaos: An Interdisciplinary Journal of Nonlinear Science}\
  }\textbf {\bibinfo {volume} {26}},\ \bibinfo {pages} {093117} (\bibinfo
  {year} {2016})}\BibitemShut {NoStop}%
\bibitem [{\citenamefont {Vincent}\ \emph {et~al.}(2018)\citenamefont
  {Vincent}, \citenamefont {Roy-Layinde}, \citenamefont {Popoola},
  \citenamefont {Adesina},\ and\ \citenamefont {McClintock}}]{VincentPRE18}%
  \BibitemOpen
  \bibfield  {author} {\bibinfo {author} {\bibfnamefont {U.~E.}\ \bibnamefont
  {Vincent}}, \bibinfo {author} {\bibfnamefont {T.~O.}\ \bibnamefont
  {Roy-Layinde}}, \bibinfo {author} {\bibfnamefont {O.~O.}\ \bibnamefont
  {Popoola}}, \bibinfo {author} {\bibfnamefont {P.~O.}\ \bibnamefont
  {Adesina}}, \ and\ \bibinfo {author} {\bibfnamefont {P.~V.~E.}\ \bibnamefont
  {McClintock}},\ }\href {\doibase 10.1103/physreve.98.062203} {\bibfield
  {journal} {\bibinfo  {journal} {Phys. Rev. E}\ }\textbf {\bibinfo {volume}
  {98}} (\bibinfo {year} {2018}),\ 10.1103/physreve.98.062203}\BibitemShut
  {NoStop}%
\bibitem [{\citenamefont {Rajamani}\ \emph {et~al.}(2014)\citenamefont
  {Rajamani}, \citenamefont {Rajasekar},\ and\ \citenamefont
  {Sanju{\'{a}}n}}]{RajamaniCiNSaNS14}%
  \BibitemOpen
  \bibfield  {author} {\bibinfo {author} {\bibfnamefont {S.}~\bibnamefont
  {Rajamani}}, \bibinfo {author} {\bibfnamefont {S.}~\bibnamefont {Rajasekar}},
  \ and\ \bibinfo {author} {\bibfnamefont {M.}~\bibnamefont {Sanju{\'{a}}n}},\
  }\href {\doibase 10.1016/j.cnsns.2014.04.006} {\bibfield  {journal} {\bibinfo
   {journal} {Commun. Nonlinear Sci. Numer. Simul.}\ }\textbf {\bibinfo
  {volume} {19}},\ \bibinfo {pages} {4003} (\bibinfo {year}
  {2014})}\BibitemShut {NoStop}%
\bibitem [{\citenamefont {Sanders}\ \emph {et~al.}(2007)\citenamefont
  {Sanders}, \citenamefont {Verhulst},\ and\ \citenamefont
  {Murdock}}]{Sanders07}%
  \BibitemOpen
  \bibfield  {author} {\bibinfo {author} {\bibfnamefont {J.~A.}\ \bibnamefont
  {Sanders}}, \bibinfo {author} {\bibfnamefont {F.}~\bibnamefont {Verhulst}}, \
  and\ \bibinfo {author} {\bibfnamefont {J.}~\bibnamefont {Murdock}},\ }\href
  {\doibase 10.1007/978-0-387-48918-6} {\emph {\bibinfo {title} {Averaging
  Methods in Nonlinear Dynamical Systems}}}\ (\bibinfo  {publisher} {Springer
  New York},\ \bibinfo {year} {2007})\BibitemShut {NoStop}%
\bibitem [{\citenamefont {Chizhevsky}(2008)}]{ChizhevskyIJBC08}%
  \BibitemOpen
  \bibfield  {author} {\bibinfo {author} {\bibfnamefont {V.~N.}\ \bibnamefont
  {Chizhevsky}},\ }\href {\doibase 10.1142/S021812740802135X} {\bibfield
  {journal} {\bibinfo  {journal} {Int. J. Bifurcation Chaos}\ }\textbf
  {\bibinfo {volume} {18}},\ \bibinfo {pages} {1767} (\bibinfo {year}
  {2008})},\ \Eprint
  {http://arxiv.org/abs/http://www.worldscientific.com/doi/pdf/10.1142/S021812740802135X}
  {http://www.worldscientific.com/doi/pdf/10.1142/S021812740802135X}
  \BibitemShut {NoStop}%
\bibitem [{\citenamefont {Ren}\ \emph {et~al.}(2017)\citenamefont {Ren},
  \citenamefont {Pan}, \citenamefont {Duan}, \citenamefont {Chapeau-Blondeau},\
  and\ \citenamefont {Abbott}}]{RenPRE17}%
  \BibitemOpen
  \bibfield  {author} {\bibinfo {author} {\bibfnamefont {Y.}~\bibnamefont
  {Ren}}, \bibinfo {author} {\bibfnamefont {Y.}~\bibnamefont {Pan}}, \bibinfo
  {author} {\bibfnamefont {F.}~\bibnamefont {Duan}}, \bibinfo {author}
  {\bibfnamefont {F.}~\bibnamefont {Chapeau-Blondeau}}, \ and\ \bibinfo
  {author} {\bibfnamefont {D.}~\bibnamefont {Abbott}},\ }\href {\doibase
  10.1103/physreve.96.022141} {\bibfield  {journal} {\bibinfo  {journal} {Phys.
  Rev. E}\ }\textbf {\bibinfo {volume} {96}} (\bibinfo {year} {2017}),\
  10.1103/physreve.96.022141}\BibitemShut {NoStop}%
\bibitem [{\citenamefont {Karle}\ \emph {et~al.}(2010)\citenamefont {Karle},
  \citenamefont {Halioua}, \citenamefont {Raineri}, \citenamefont {Monnier},
  \citenamefont {Braive}, \citenamefont {Le~Gratiet}, \citenamefont {Beaudoin},
  \citenamefont {Sagnes}, \citenamefont {Roelkens}, \citenamefont {van Laere},
  \citenamefont {Van~Thourhout},\ and\ \citenamefont {Raj}}]{KarleAPL10}%
  \BibitemOpen
  \bibfield  {author} {\bibinfo {author} {\bibfnamefont {T.~J.}\ \bibnamefont
  {Karle}}, \bibinfo {author} {\bibfnamefont {Y.}~\bibnamefont {Halioua}},
  \bibinfo {author} {\bibfnamefont {F.}~\bibnamefont {Raineri}}, \bibinfo
  {author} {\bibfnamefont {P.}~\bibnamefont {Monnier}}, \bibinfo {author}
  {\bibfnamefont {R.}~\bibnamefont {Braive}}, \bibinfo {author} {\bibfnamefont
  {L.}~\bibnamefont {Le~Gratiet}}, \bibinfo {author} {\bibfnamefont
  {G.}~\bibnamefont {Beaudoin}}, \bibinfo {author} {\bibfnamefont
  {I.}~\bibnamefont {Sagnes}}, \bibinfo {author} {\bibfnamefont
  {G.}~\bibnamefont {Roelkens}}, \bibinfo {author} {\bibfnamefont
  {F.}~\bibnamefont {van Laere}}, \bibinfo {author} {\bibfnamefont
  {D.}~\bibnamefont {Van~Thourhout}}, \ and\ \bibinfo {author} {\bibfnamefont
  {R.}~\bibnamefont {Raj}},\ }\href {\doibase 10.1063/1.3319667} {\bibfield
  {journal} {\bibinfo  {journal} {Journal of Applied Physics}\ }\textbf
  {\bibinfo {volume} {107}},\ \bibinfo {pages} {063103} (\bibinfo {year}
  {2010})},\ \Eprint {http://arxiv.org/abs/https://doi.org/10.1063/1.3319667}
  {https://doi.org/10.1063/1.3319667} \BibitemShut {NoStop}%
\bibitem [{\citenamefont {Makles}\ \emph {et~al.}(2015)\citenamefont {Makles},
  \citenamefont {Antoni}, \citenamefont {Kuhn}, \citenamefont {Del\'{e}glise},
  \citenamefont {Briant}, \citenamefont {Cohadon}, \citenamefont {Braive},
  \citenamefont {Beaudoin}, \citenamefont {Pinard}, \citenamefont {Michel},
  \citenamefont {Dolique}, \citenamefont {Flaminio}, \citenamefont {Cagnoli},
  \citenamefont {Robert-Philip},\ and\ \citenamefont {Heidmann}}]{MaklesOL15}%
  \BibitemOpen
  \bibfield  {author} {\bibinfo {author} {\bibfnamefont {K.}~\bibnamefont
  {Makles}}, \bibinfo {author} {\bibfnamefont {T.}~\bibnamefont {Antoni}},
  \bibinfo {author} {\bibfnamefont {A.~G.}\ \bibnamefont {Kuhn}}, \bibinfo
  {author} {\bibfnamefont {S.}~\bibnamefont {Del\'{e}glise}}, \bibinfo {author}
  {\bibfnamefont {T.}~\bibnamefont {Briant}}, \bibinfo {author} {\bibfnamefont
  {P.-F.}\ \bibnamefont {Cohadon}}, \bibinfo {author} {\bibfnamefont
  {R.}~\bibnamefont {Braive}}, \bibinfo {author} {\bibfnamefont
  {G.}~\bibnamefont {Beaudoin}}, \bibinfo {author} {\bibfnamefont
  {L.}~\bibnamefont {Pinard}}, \bibinfo {author} {\bibfnamefont
  {C.}~\bibnamefont {Michel}}, \bibinfo {author} {\bibfnamefont
  {V.}~\bibnamefont {Dolique}}, \bibinfo {author} {\bibfnamefont
  {R.}~\bibnamefont {Flaminio}}, \bibinfo {author} {\bibfnamefont
  {G.}~\bibnamefont {Cagnoli}}, \bibinfo {author} {\bibfnamefont
  {I.}~\bibnamefont {Robert-Philip}}, \ and\ \bibinfo {author} {\bibfnamefont
  {A.}~\bibnamefont {Heidmann}},\ }\href {\doibase 10.1364/OL.40.000174}
  {\bibfield  {journal} {\bibinfo  {journal} {Opt. Lett.}\ }\textbf {\bibinfo
  {volume} {40}},\ \bibinfo {pages} {174} (\bibinfo {year} {2015})}\BibitemShut
  {NoStop}%
\bibitem [{\citenamefont {{Bogoliubov}}\ and\ \citenamefont
  {{Mitropolski}}(1961)}]{Bogoliubov61}%
  \BibitemOpen
  \bibfield  {author} {\bibinfo {author} {\bibfnamefont {N.~N.}\ \bibnamefont
  {{Bogoliubov}}}\ and\ \bibinfo {author} {\bibfnamefont {Y.~A.}\ \bibnamefont
  {{Mitropolski}}},\ }\href@noop {} {\emph {\bibinfo {title} {{Asymptotic
  Methods in the Theory of Non-Linear Oscillations}}}}\ (\bibinfo  {publisher}
  {New York: Gordon and Breach},\ \bibinfo {year} {1961})\BibitemShut {NoStop}%
\end{thebibliography}%

\end{document}